\documentclass[12pt]{iopart}

\usepackage{setspace}
\usepackage{graphicx}
\usepackage{dcolumn}
\usepackage{bm}
\usepackage{iopams}

\begin{document}

\title{Modified percolation theory and its relevance to quantum critical phenomena}

\author{Tom Heitmann$^1$, John Gaddy$^2$, and Wouter Montfrooij$^{1,2}$}

\address{$^1$Missouri Research Reactor, University of Missouri, Columbia, MO 65211 USA}
\address{$^2$Department of Physics and Astronomy, University of Missouri, Columbia, MO 65211 USA}

\date{14 November 2012}

\begin{abstract}
We present a new family of percolation models. We show, using theory and computer simulations, that this class represents a new universality class. Interestingly, systems in this class appear to violate the Harris criterion, making model systems within this class ideal systems for studying the influence of disorder on critical behavior. We argue that such percolative systems have already been realized in practice in strongly correlated electron systems that have been driven to the quantum critical point by means of chemical substitution.
\end{abstract}

\maketitle
\section{Introduction}
Percolation theory and its many applications to physical phenomena does not require a lengthy introduction \cite{stauffer,sahimi,orbach}. Percolation theory describes how the response of a system changes upon removal of its elements or of the connections between these constituents. Once enough elements have been removed and the percolation threshold is approached, the system displays critical behavior characterized by universal exponents and scaling functions. The percolation threshold itself-- that is, the point where there is the flimsiest of  connections between one side of the system and the other following a fractal path of surviving elements-- depends both on the dimensionality of the system and on the number of neighbors in the undiluted lattice\cite{stauffer}. Many critical phenomena have now been mapped\cite{sahimi} onto a percolation problem.\\

In this paper we investigate the behavior of systems where the removal of its constituents has been restricted in the sense that we only remove elements from the cluster that spans the width of the system. Normally, elements can be removed from all occupied sites, not merely from the cluster that spans the lattice. We show that such a restriction not only shifts the percolation threshold, it also changes the powerlaws that characterize the critical behavior of the system. An example of such a powerlaw is how the infinite cluster $P$ loses elements in the vicinity of the percolation threshold $p_c$: $P(x) \sim x^{\beta}$ with $x = p-p_c$ and $p$ the fraction of occupied (or surviving) lattice sites.\\
An important consequence of the above restriction is that the number of clusters present in the system upon approaching the percolation threshold no longer increases exponentially when compared to the unrestricted percolation scenario. While the number of isolated clusters still increases because more and more sites are removed from the lattice spanning cluster, allowing for smaller clusters to peel off, the ever increasing numbers of isolated clusters are not allowed to subdivide further. This changes the critical powerlaw behavior present in standard percolative systems into analytic behavior in our restricted case.\\

When the number of isolated clusters no longer diverges upon approaching the percolation threshold, then the Harris criterion \cite{harris} is violated. Whenever the criterion is satisfied, then we can expect that impurities will not end up determining the critical behavior of a system. In contrast, when the criterion is violated, then impurities cannot be disregarded. As such, our restricted percolation model should prove to be an ideal model system for studying the influence of impurities on critical behavior.\\

We detail our modified percolation model in the next section where we will derive the relationship (between the restricted and unrestricted cases) for the critical exponent governing the demise of the lattice spanning cluster upon approaching $p_c$. We also present the results of Monte Carlo computer simulations in that section to support our claims. In the last section of this paper we show that the onset of magnetic order in chemically substituted Kondo lattice systems represents a practical realization of a restricted percolative system. This identification, together with its implicit violation of the Harris criterion, opens the door for a much better understanding of the ordering tendencies of chemically doped quantum critical systems whose response has, thus far \cite{stewart}, eluded theoretical description.

\section{Theory and computer simulations for the restricted percolation model}
In standard percolation, we start off with a fully occupied lattice, and then we remove lattice sites at random. When more and more sites are removed, a situation is reached where individual sites or groups of sites (clusters) become isolated from the main group of sites that spans the lattice, the so-called 'lattice spanning' or 'infinite' cluster. Upon removal of more sites, we will reach the point where the infinite cluster breaks up. This point is called the percolation threshold, and denoted by $p_c$; the critical concentration $p$ of lattice sites below which we cannot find a cluster that connects one side of the lattice to the opposite side. The percolation threshold depends on the dimensionality of the system, as well as on the number of neighbors $Z$ that any site is connected to.\\
The behavior of certain quantities, such as the number of occupied sites $P(p)$ that are part of the infinite cluster or the number of isolated clusters in the system $M(p)$, displays a powerlaw dependence when measured as a function of how far the system is removed from the percolation threshold. For instance, the strength of the infinite cluster close to the percolation threshold can be expressed as
\begin{equation}
P(p)=P_0(p-p_c)^{\beta}+(p-p_c) $ for $ p-p_c >0.
\label{infcluster}
\end{equation}
The argument $p-p_c$ is a measure of how far the system is removed from the percolation threshold, and the exponent $\beta$ is called the critical exponent. The critical exponents depend on the dimensionality of the system, but not on the underlying details such as the number of nearest neighbors. Given the unimportance of such details, these exponents are referred to as universal exponents, and we lump groups of systems together into universality classes characterized by a particular set of exponents.\\

When we restrict the removal of sites to only allow sites to be removed from the infinite cluster, we obviously change the percolation threshold. After all, when we only remove sites from the infinite cluster, every site removal will serve to weaken the infinite cluster and we should reach the percolation threshold quicker, meaning that $p_{c,restricted}>p_{c,unrestricted}$. However, this shift in percolation threshold by itself does not necessarily imply a change in universal exponent. Nonetheless, the following straightforward argument demonstrates that the restricted model should have different critical exponents, and, therefore, that we will have a different universality class.\\

\begin{figure}[t]
\begin{center}
\includegraphics[width=380pt]{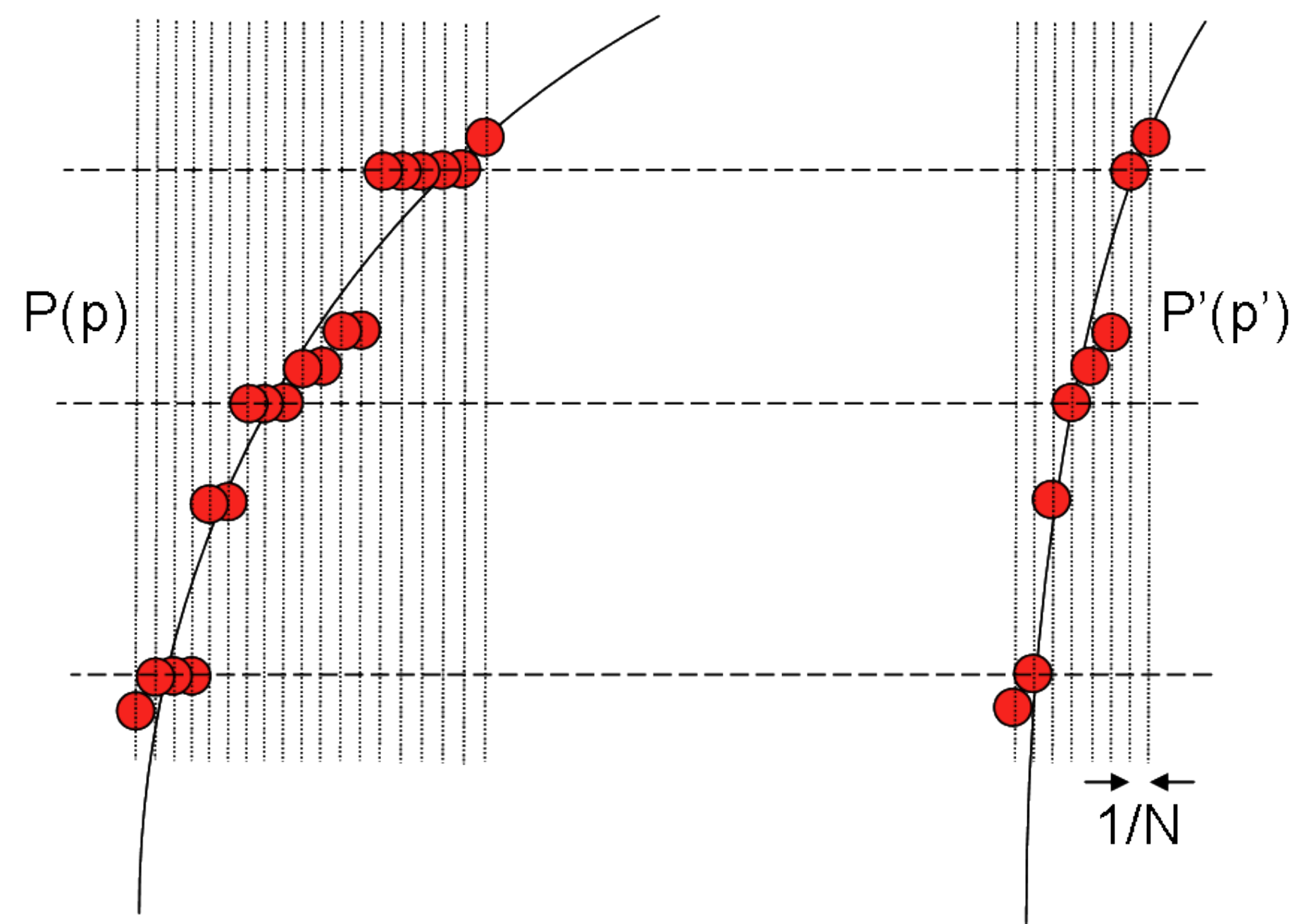}
\caption{A schematic comparison between the demise of the infinite cluster in the restricted model (on the right) and in the standard percolation model (on the left). These snapshots show the change in infinite cluster strength when the lattice occupancy $p$ is changed by $1/N$ (vertical lines) at every site removal. The dashed horizontal lines signify identical morphologies for $P'(p')$ and $P(p)$. For standard percolation (left), there can be changes in occupancy $p$ that do not result in any change in membership of the infinite cluster; such non-changes occur when sites are removed from isolated clusters. On average, it takes $p/P$ steps to remove a site from the infinite cluster in standard percolation versus only one step in restricted percolation. This can be seen to result in a change of the slope of the powerlaw that describes the evolution of the infinite clusters (black solid lines). The situation shown here-- where it takes 18 steps to accomplish in the standard percolation scenario what only takes 7 steps to accomplish in the restricted percolation scenario-- would correspond to $P \sim 0.1$ when $p \sim 0.25$.}\label{pintro}
\end{center}
\end{figure}

In standard percolation when a fraction of sites $p$ are still occupied, while a fraction $P(p)$ sites form the infinite cluster, on average it will take $p/P(p)$ site removals to remove just one site from the infinite cluster. In our restricted percolation scenario, it requires only one site removal to remove a site from the infinite cluster. Close to the percolation threshold, the strength of the infinite cluster $P(p)$ will rapidly diminish, necessitating increasingly more site removals in the unrestricted scenario before the infinite cluster is affected again. It is this 'increasingly more' part, captured by the $p/P(p)$ dependence, that changes the critical exponent. This is shown in Fig. \ref{pintro}.\\

In addition, we can argue that the number of isolated clusters does not display a powerlaw behavior close to the percolation threshold under the restricted scenario. In standard percolation, clusters peel from off the infinite cluster. However, these clusters are allowed to break up further, increasing the total number of clusters. The
closer one gets to the percolation threshold, the more and more sites will be removed
from finite clusters as opposed to the infinite cluster, simply because sites slated for removal are picked
at random, and there is much more mass in all the finite clusters combined than in the
infinite cluster. Therefore, not only does the number of clusters increase the closer
we get to the threshold, this number increases increasingly faster because the mass
of the infinite cluster gets less and less. This explains the divergent behavior of the
number of clusters in the standard percolation model.\\

Conversely, in our restricted case, we only remove sites from the infinite cluster.
Once we get close to the percolation threshold, the number of finite clusters we add
is simply given by the number of red bonds we break on the infinite cluster. Red bonds \cite{stauffer} are the sites in the infinite cluster network at which two parts of the infinite cluster are only connected through a single site. Perhaps
this red bond breaking happens once in every four bonds we break. But whatever the ratio is in reality,
it does not depend critically on how close we are to the threshold, and, therefore, the
number of finite clusters that peel off should be an analytic function since the speed at
which finite clusters peel off does not increase increasingly faster.

\subsection{Critical exponents in restricted percolation}
We can capture the preceding conceptual reasoning in equations, which will provide us with a relationship between the critical exponent $\beta$ in the restricted and unrestricted cases. In order to distinguish between the two scenarios while keeping the number of subscripts limited, we attach a prime to the quantities in the restricted scenario. For instance, $\beta'$ refers to the critical exponent for the infinite cluster in the restricted case.\\

In order to setup our equations, we imagine a computer simulation where we start with a fully occupied lattice. We assign a number to every lattice site, and we roll dice to see which of these numbers is chosen for site removal. At first, we will not notice any distinction between the restricted and unrestricted cases as many sites need to be removed before isolated clusters form. But when more and more isolated clusters have formed, we will notice that sometimes we have to roll the dice again in the restricted case since a number came up that belongs to an occupied site in an isolated cluster. Of course, in the unrestricted case we will simply remove this site, thereby lowering $p$, before rolling again.\\

This imagery demonstrates two things. First, the percolation threshold will be reached at a lower value of $p$ in the unrestricted case versus the restricted case. Second, the shape and size of the infinite cluster will go through identical points in both scenarios, the only difference being the value of $p$ at which these identities occur (Fig. \ref{pintro}). Thus, when we roll the dice this way, we will always be able to find a solution to the following equation for any value of $p$ with $p>p_c$:
\begin{equation}
P'(p')=P(p).\\
\label{mormor}
\end{equation}
Since this equation holds for all values of $p$, we can also relate the slope of $P'(p')$ to that of $P(p)$. In fact, the slope of $P'(p')$ will be steeper by a factor of $p/P(p)$ than the slope of $P(p)$ simply because it takes $p/P(p)$ steps in the unrestricted case to achieve the identical demise of the infinite cluster which is achieved in one step in the restricted case. This is shown in Fig. \ref{pintro} and it is captured in the following equation (with $x=p-p_c$ and $x'=p'-p'_c$):
\begin{equation}
\frac{d P'(x')}{d x'}= \frac{d P(x)}{d x} \frac {p}{P(x)} $ for $ P'(x')=P(x).\\
\label{slopes}
\end{equation}
Using eqn \ref{infcluster} to describe the critical behavior of the infinite cluster in the unrestricted case, and the similar expression for the restricted case
\begin{equation}
P'(p')=P'_0(p'-p'_c)^{\beta'}+(p'-p'_c) $ for $ p'-p'_c >0,
\label{infcluster2}
\end{equation}
we can rewrite eqn \ref{slopes} as follows:
\begin{equation}
P'_0 \beta' x'^{\beta'-1}=P_0 \beta x^{\beta-1}\frac {p}{P_0 x^{\beta}}=\frac{\beta (x+p_c)}{x} \approx \frac{\beta p_c}{x}.
\label{slopes2}
\end{equation}
In the above equation, we ignored the terms $(p-p_c)$ and $(p'-p'_c)$ close to the percolation thresholds where the powerlaw terms dominate. In this approximation, the solution to eqn \ref{mormor} is given by
\begin{equation}
x= (\frac{P'_0}{P_0})^{1/\beta}(x')^{\beta'/\beta}.
\label{solmormor}
\end{equation}
Substituting eqn \ref{solmormor} into eqn \ref{slopes2} we find
\begin{equation}
\frac{\beta p_c}{x} =\frac{\beta p_c}{(P'_0/P_0)^{1/\beta}(x')^{\beta'/\beta}}= P'_0 \beta' x'^{\beta'-1}.
\label{slopes3}
\end{equation}
Since this equation holds for all $x'$, we must have that both the prefactors, as well as the powers of $x'$ match on either side of the equation. Rearranging the terms we find
\begin{eqnarray}
\begin{array}{lll}
\beta' &=&{\beta}/({1+\beta})\\
P'_0 &=& (P_0)^{1/(1+\beta)}((1+\beta)p_c)^{\beta/(1+\beta)}.\\
\end{array}
\label{eatthat}
\end{eqnarray}
These equations readily capture that the restricted percolation model is in a different universality class. They also show that we can use the literature values for the unrestricted model to compare to our restricted case. Bearing in mind that it takes ($p/P-1)$ more steps in the unrestricted case to achieve the same demise that the infinite cluster suffers in one step in the restricted case, we can also express the shift in percolation threshold $\Delta p_c$ between the two cases as
\begin{equation}
\Delta p_c=\int_{p'_c}^{1}\left[ \frac{p}{P(p)}-1 \right ]dp'.
\label{pshift}
\end{equation}
The integral is to be evaluated at all points for which $P'(p')=P(p)$, something which cannot be easily done without prior knowledge of $P'(p')$, but which can be used as a check in computer simulations that calculate both $P(p)$ and $P'(p')$ at the same time.\\

The above equations are exact, however, we cannot calculate the other critical exponents with the same rigor. Critical exponents (and observable physical quantities) are related to the moments of the cluster distribution $\sum s^k n_s(p')$, with $n_s(p')$ the number of clusters per site that have $s$ members while the summation runs over all finite sized clusters. For instance, when $k = 1$, then we can use the following sum rule to relate the exponent $\beta$ to the first moment of the $n_s(p')$ distribution:
\begin{equation}
P'(p')+\sum_s sn_s(p')=p'.\\
\end{equation}

We can still say something meaningful about the other moments when we make the following simplification. Whenever sites become isolated from the infinite cluster, then we will have in most cases that these isolated sites peel off as one cluster. Sometimes when a site is removed, it will give rise to more than one cluster peeling off provided these clusters were all linked to the infinite cluster through the same site. Such events are far less common than the peeling off of just one cluster, but they do occur. When we assume that all sites that peel off from the infinite cluster end up in one single isolated cluster, then we can calculate the moments of the cluster distribution function $n_s(p')$. However, we do not know if we still capture the true critical behavior under this assumption. It is clear that we would underestimate the overall number of isolated clusters that we end up with ($k = 0$), and that we would overestimate all moments for $k > 1$, but it is unclear if this would affect the critical exponents.\\

We derive the following equation for the moments in Appendix A, using the above simplification. We find that the non-analytic contributions to the moments are given by
\begin{equation}
\sum_s s^{k} n_s(p') \sim (p'-p_c')^{1-k+k\beta'}.
\label{moments}
\end{equation}
This equation is consistent with our earlier reasoning that the number of isolated clusters, as given by the $k=0$ moment of the above equation, does not diverge at the percolation threshold, and is in fact an analytical function of $p'-p'_c$. This assertion is also borne out by computer simulations that we present later in this paper.
\subsection{The Harris criterion}
The preceding discussion on the critical exponents bring us to the discussion of the so-called Harris criterion.
From the beginning of the study into critical phenomena it was known that an ideal
system (as described in standard percolation theory) would undergo critical scaling \cite{stauffer,sahimi} as the
system approached the percolation threshold. The thought initially was that a given
concentration of impurities (or defects) would be enough to destroy the fragile scaling
that was observed near the critical point: as the system approached
criticality, the defects would force the system to divide up into fragmented
sections that would undergo a phase transition at different temperatures. Because
different parts of the system would undergo this phase transition at different temperatures, the critical
scaling would be lost as the divergences of correlation length, specific heat, and other critical
quantities were spread out over a broad temperature scale. As described above, this
would then mean that any scaling would be unrecognizable as the power law would be
smeared out.\\

However, after careful study, it was discovered that in the presence of small
amounts of disorder \cite{harris} the system still retains its sharp scaling behavior on approach to phase transition, provided that the system satisfied the requirement that
$\nu > 2/d$, with $\nu$ the critical exponent describing the growth of the average diameter of isolated clusters and $d$ the dimensionality of the system. Thus, if we have a system with a given amount of disorder (through the addition
of defects or impurities), we are able to deduce whether or not this disorder will
lead to a system with a smeared or with a clean critical point and accompanying power law dependences.\\

For our restricted percolation scenario, we find that the Harris criterion is violated. This can most easily be seen using the scaling relationship \cite{orbach} between the critical exponent $\alpha$ for the number of isolated clusters and the exponent $\nu$ describing the growth in correlation:
\begin{equation}
\alpha= 2-2d \nu.
\end{equation}
We can see that the Harris criterion corresponds to $\alpha < 0$. Since we find analytical behavior for the number of isolated clusters ($\alpha = 1$), we find $\nu = 1/d$. However, we have to bear in mind that the value of $\alpha = 1$ relied on an assumption, so we should allow for the possibility that $\alpha$ equals zero. When $\alpha = 0$, then $\nu = 2/d$, still in violation of the Harris criterion but only marginally so.\\

Violation of the Harris criterion by itself in not enough to conclude that disorder will determine the critical properties close to the threshold. In fact, in the case where
the Harris criterion is violated, Chayes {\it et al.} \cite{chayes} showed that a new critical point will
arise in which the conventional power law scaling exists but with new exponents that
satisfy the Harris criterion. Should this be the case, then our restricted percolation model would be an ideal system to actually study these changes in scaling behavior as predicted by Chayes {\it et al.}.\\

We used the word 'should' in the preceding sentence because the proof by Chayes {\it et al.} was based on the assumption that there is no relevant length scale smaller
than the correlation length close to the percolation threshold. In our restricted case, the clusters do
not change their morphology once they peel off, and, therefore, we have-- for any value of the average correlation length-- clusters
that will be smaller than this length scale, and clusters that will be larger than this
length scale. As such, the chief assumption \cite{chayes} of Chayes {\it et al.} may no longer be valid. In either case, whether the proof stands up or not in our restricted case, the restricted percolation model can be regarded as an ideal set of model systems to study the influence of disorder on critical behavior.

\subsection{Computer simulations}
In order to check the predictions pertaining to our restricted model we have performed
Monte Carlo computer simulations. We have opted to do our simulations for the three dimensional body centered lattice. This is the lattice that pertains to one heavily studied family of quantum critical systems\cite{stewart}, the so-called 122-compounds that crystalize in the ThCr$_2$Si$_2$ structure.
Our simulations are limited in accuracy as we only simulated lattices of about 100,000 lattice sites in three dimensions and because the restricted model is valid only above the percolation threshold, resulting in a rather small region of critical behavior as opposed to models where the threshold can also be approached from below.\\

\begin{figure}[t]
\begin{center}
\includegraphics[width=380pt]{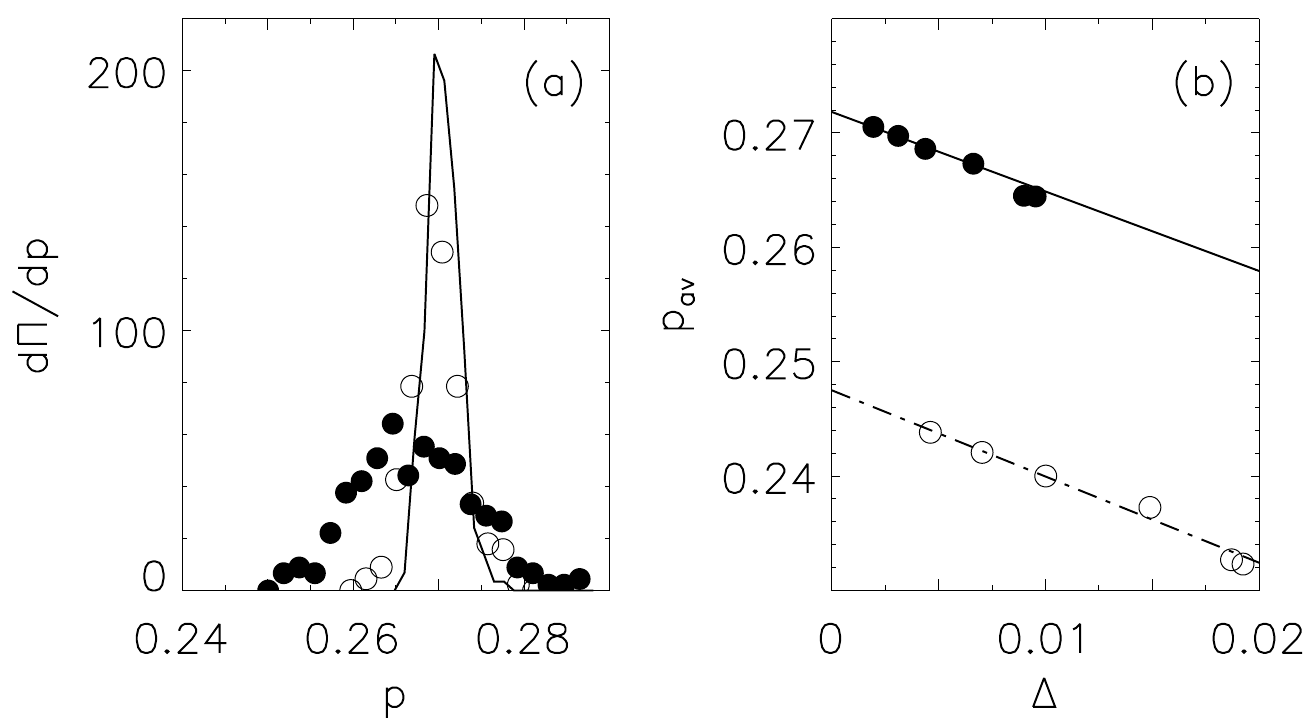}
\caption{(a) The distribution of thresholds $d\Pi/dp$ ($dR/dp$ in Stauffer\cite{stauffer}) as obtained from 250 runs on a lattice of 15$^3$ unit cells (filled symbols), 25$^3$ (open symbols) and 35$^3$ (solid line). The distribution is characterized by an average $p_{av}$ and a width $\Delta$. With increasing system size the distribution can be seen to narrow while approaching the true threshold (0.272 in this case). (b) Using the relationship \cite{stauffer} between the average percolation threshold $p_{av}$ and the width $\Delta$ of the distribution ($p_{av}-p_c \sim \Delta$) we obtain $p_c$ and $p_c'$ by extrapolating $\Delta \rightarrow$ 0. We find $p_c'$=0.2719 $\pm$ 0.0005 (filled symbols) and $p_c$=0.2475 $\pm$ 0.001 (open symbols).\label{threshold}}
\end{center}
\end{figure}

We have performed our simulations for lattices of $L^3$ unit cells, with $L$= 10, 12, 15, 20, 25 and 35, using periodic boundary conditions. We average the results for each lattice size over 250 runs. We use the smaller sized simulations to be able to more accurately determine the percolation threshold, and we use the largest simulation size to test the predictions for our restricted model. For each run we determine the percolation threshold as the point where the infinite cluster breaks up. In this way, we obtain a distribution of percolation thresholds for each lattice size. Using the Levinshtein method\cite{levinshtein}, we can then estimate the true percolation threshold-- free from finite size simulation effects-- based upon the width $\Delta$ and average $p_{av}$ of the threshold distribution. This is shown in Fig. \ref{threshold}, yielding $p_c$=0.2475 $\pm$ 0.001 and $p_c'$=0.2719 $\pm$ 0.0005. The literature value \cite{sahimi} for the unrestricted case is $p_c$=0.2465, while the shift in threshold values between the two scenarios is borne out by evaluating the integral of eqn \ref{pshift} using the numerical values for $P(p$); this procedure predicts a shift of 0.023$\pm$ 0.001, in reasonable agreement with the observed shift of 0.024.\\

\begin{figure}[b]
\begin{center}
\includegraphics[width=380pt]{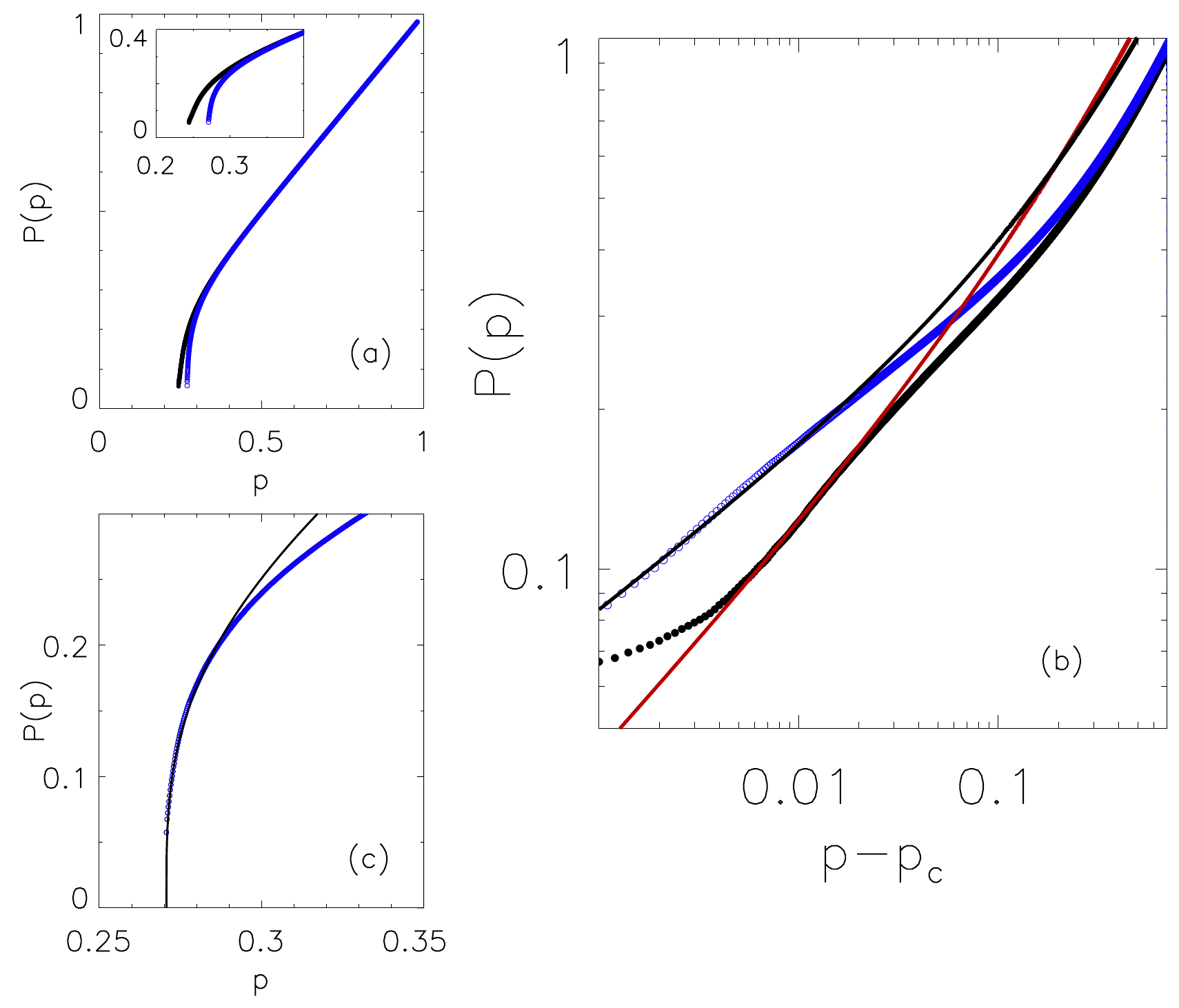}
\caption{Simulation results based on the average of 250 runs performed on 35x35x35 body centered unit cells. (a and inset) The probability $P(p)$ that an occupied site belongs to the infinite cluster, both for the standard percolation model (black symbols) and for our restricted model (blue symbols). The difference between the two models results in a clearly identifiable shift in percolation threshold. (b) When plotted on a log-log scale, the critical behavior of $P(p)$ can be seen to follow a power law dependence. The unrestricted model (black symbols) has been fitted to eqn \ref{infcluster}, yielding the red curve. Using the fitted values in conjunction with eqn \ref{eatthat} yields the black curve that follows the restricted model data (blue symbols) quite well. Note that both cases yield a critical region that spans the same range in $P(p)$ on the vertical axis. (c) Same as panel (b) except that now only the restricted case has been shown as a function of $p$. The black curve through the data contains no adjustable parameters, the percolation threshold has been taken from Fig. \ref{threshold}.\label{ppp}}
\end{center}
\end{figure}

Next, we verify the prediction that the restricted percolation model represents a new universality class with an exponent $\beta'=\beta/(\beta +1)$ describing the critical behavior of the strength of the infinite cluster. We show this verification in Fig. \ref{ppp}. In order to test our prediction for the critical exponent, we first fitted the unrestricted case to eqn \ref{infcluster} with $\beta= 0.41$ taken from the literature \cite{sahimi} and $P_0$ as the only free parameter. This one parameter fit is displayed in Fig. \ref{ppp}b.\\

Using the literature value $\beta= 0.41$ and the fitted value for $P_0$, we then use eqn \ref{eatthat} to calculate $\beta'$ and $P'_0$. We then use these values to describe $P'(p')$. As can be seen in Fig. \ref{ppp}b and \ref{ppp}c, these values-- when plugged into eqn \ref{infcluster2}-- yield a very good description of $P'(p')$. Note that these powerlaws for the restricted case do not contain any adjustable parameters. Thus, our computer simulations verify our assertion that the restricted percolation model constitutes a new universality class with the critical exponent describing the infinite cluster given by eqn \ref{eatthat}.\\

\begin{figure}[t]
\begin{center}
\includegraphics[width=380pt]{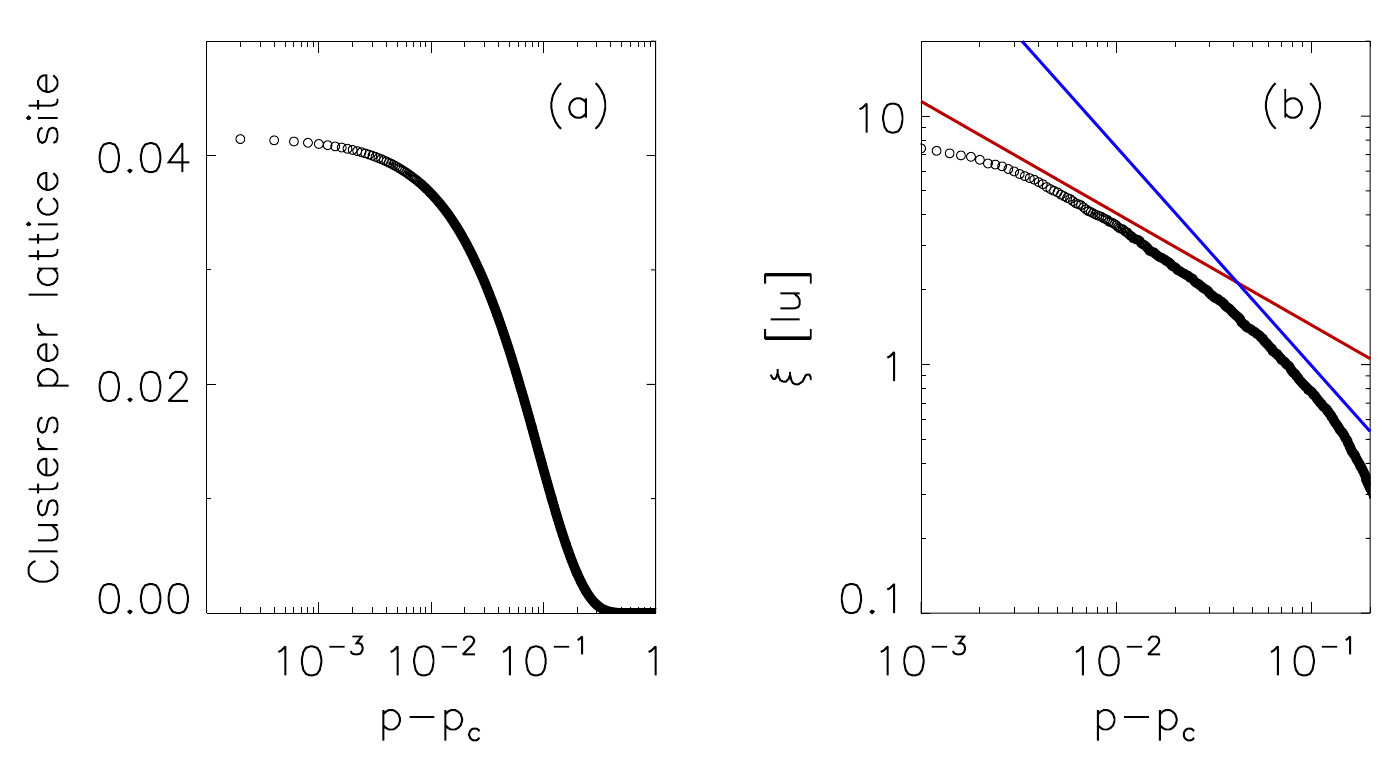}
\caption{Simulation results based on the average of 250 runs performed on 35x35x35 body centered unit cells. (a) The number of clusters per lattice site is shown on a linear-log scale plot. There is no evidence of any powerlaw behavior. Note that the horizontal scale has been extended to $10^{-4}$; the range $10^{-3}$-$10^{-4}$ is susceptible to errors in the determination of the percolation threshold, however, it can be seen that even if such errors are present that they will not introduce any powerlaw dependence. (b) The correlation length $\xi$ does display critical behavior in this log-log plot, but because of transition rounding errors, it is unclear what the actual value for the powerlaw would be. The blue line is given by the power law dependence \cite{sahimi} for the unrestricted percolation scenario $\xi \sim (p'-p'_c)^{-0.88}$, the red line is given by $\xi \sim (p'-p'_c)^{-0.45}$. Violation of the Harris criterion occurs for exponents $> -2/3$. Thus, the red line corresponds to a clear violation of the Harris criterion.\label{m0}}
\end{center}
\end{figure}

In Fig. \ref{m0}a we show the number of isolated clusters present in our restricted percolation model as calculated from our computer simulations. It is clear from this figure that there is no hint of any powerlaw behavior in the evolution of the number of isolated clusters, even if we allow for some uncertainty in the exact percolation threshold and for transition rounding.\\

We have also determined the dependence of the correlation length $\xi$-- a measure of the average diameter of the isolated clusters-- on the distance to the percolation threshold. We show the results in Fig. \ref{m0}b. Compared to our results for the number of isolated clusters and for the strength of the infinite cluster, the correlation length data are less accurate. The main reason for this is the sensitivity of $\xi$ to the size of the simulation. $\xi$ depends strongly on the largest clusters present, and the size of the largest clusters is cut-off by the size of the simulation. Such transition rounding effects are already visible in Fig. \ref{m0}b. for $p'-p'_c \approx 10^{-3}$.\\

What we can conclude from the data shown in the figure, however, is that the critical exponent describing the evolution of $\xi$ is smaller than the known value for percolation phenomena in unrestricted systems, and that this exponent certainly appears to be smaller than $2/d$, the value below which the Harris criterion is violated. But as mentioned, the overall size of the simulation is too small to determine the value of the critical exponent.

\section{Relevance to chemically doped quantum critical systems}
The discovery of quantum critical systems \cite{stewart} has created a new chapter in the study of phase transitions. In quantum critical systems, a second order phase transition is suppressed down to zero Kelvin by external means such as applying pressure or magnetic fields, and the order-disorder boundary is then controlled by quantum fluctuations as opposed to thermal fluctuations. The result is that critical exponents describing such a transitions are now expected to be modified to include the dynamical scaling exponent\cite{sachdev}. This changes the universality class and moves the system up to a higher effective dimension, which in turn should diminish the influence of fluctuations, perhaps even up to the point where the system surpasses the upper critical dimension and fluctuations are only marginally important. However, despite these obvious truths, the critical exponents regarding such transitions have not been worked out\cite{stewart}. In fact, it is unclear whether universal behavior even occurs in quantum critical systems: it might well be the case that intrinsic disorder\cite{neto} ultimately determines the critical behavior.\\

The second order transition that has been studied most extensively \cite{stewart} is the magnetic ordering that takes place in Kondo lattice systems. Such systems have magnetic ions embedded into the unit cell, and upon cooling down, these magnetic ions tend to align with the neighbors. However, in these metals this ordering is opposed by the conduction electrons who have a tendency of shielding the moments through the Kondo mechanism \cite{sachdev,kondo}. Whether the system ends up in a magnetically ordered ground state or in a heavy fermion state where the conduction electrons have formed non-magnetic singlets with the magnetic ions depends critically on the strength of the interaction between the conduction electrons and the moments of the magnetic ions. This interaction strength in turn depends very sensitively on the atomic separation between the magnetic ions and the non-magnetic ions. This observation brings us closer to the relevance of percolation theory to the ordering tendencies in these unusual \cite{sachdev} metals.\\

In order to suppress the magnetic ordering transition in Kondo lattice systems to zero Kelvin-- which is done to bring to the fore the consequences of having conduction electrons being strongly coupled to the magnetic ions and to each other-- one tweaks the coupling strength between the conduction electrons and the magnetic ions. In practice, this tweaking is done by applying chemical pressure. One substitutes smaller (larger) ions for the ions already present, thereby shrinking (expanding) the lattice, resulting in a stronger (weaker) coupling. An example of this process is substituting Ru ions on Fe-sites in CeFe$_2$Ge$_2$, resulting in a system that will order magnetically once $\sim$24.6\% of the Fe-ions have been substituted \cite{brazil}. Many more examples of systems that have been driven to exhibit an order-disorder transition at 0 K can be found in the review by Stewart \cite{stewart}.\\

Chemical substitution introduces disorder, but, more importantly, it introduces a distribution of temperatures below which individual magnetic ions will be shielded by the conduction electrons. On a microscopic level in the doped system, we will find a distribution of spacings between the outer shells of neighboring magnetic and non-magnetic ions, resulting in a distribution of coupling strengths between the magnetic moments and the conduction electrons \cite{bernal}. This in turn implies that when such a substituted system is cooled down, that some magnetic ions will be shielded while others might survive down to lower temperatures, or even resist shielding altogether.\\

The consequence of this substitution is a percolative system in which magnetic moments disappear through shielding upon cooling. Which moments disappear first is determined at random since the chemical substitution will have taken place in random unit cells. If enough moments survive down to temperatures where the moments line up, then we will encounter long-range magnetic ordering. If too few survive, then long-range order will not materialize. The critical composition would be given by there still, but only barely, being an infinite cluster present at 0 K.\\

The percolation described in the preceding paragraphs is an unavoidable consequence of chemical substitution in combination with the underlying physics of the shielding process (namely, the very sensitive dependence on interatomic distances). However, what is described is standard percolation, not the restricted model that we introduced in the preceding sections. For the latter, we also need to consider the importance of finite size effects.\\

\begin{figure}[t]
\begin{center}
\includegraphics[width=380pt]{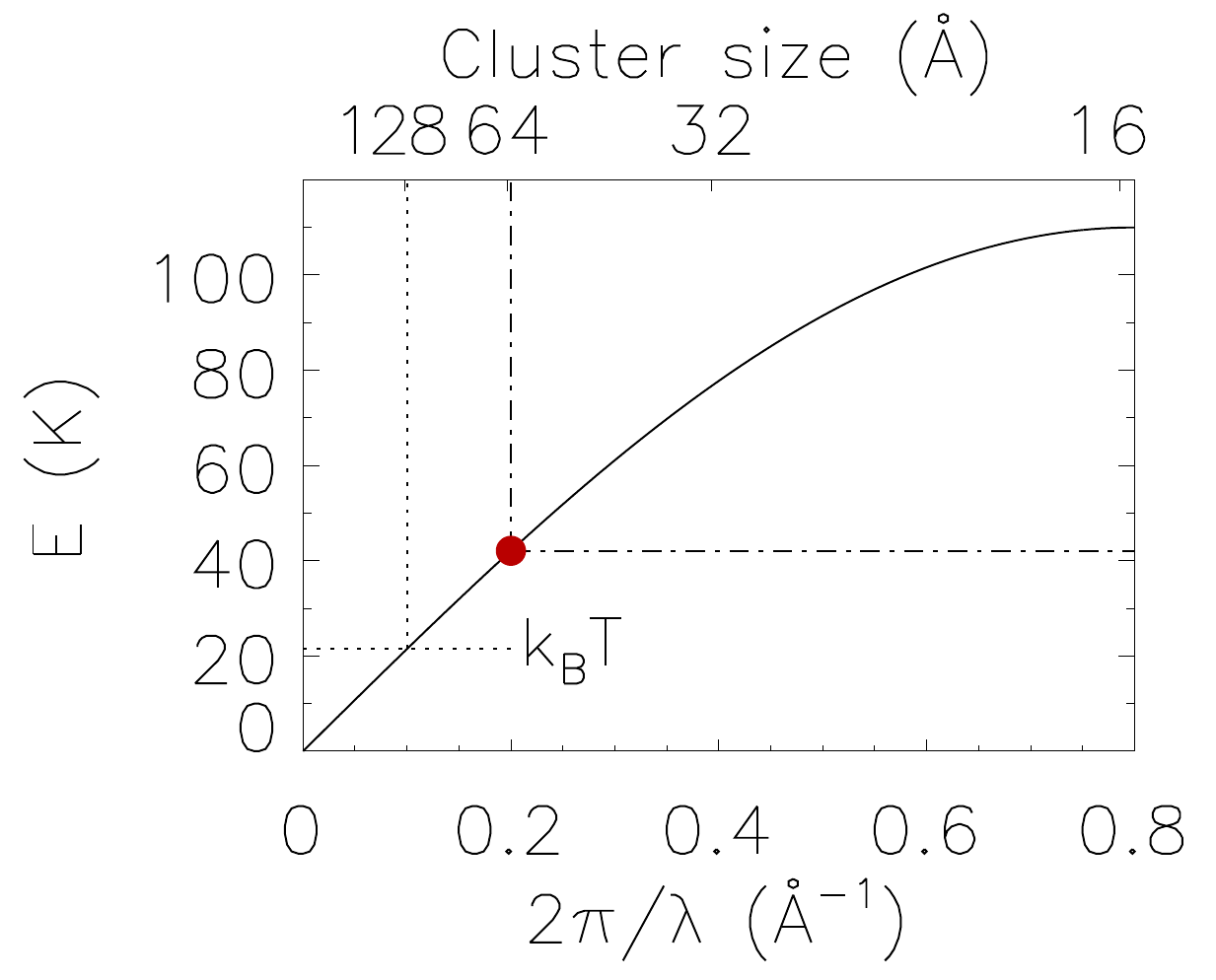}
\caption{A typical magnon (or spin wave)  dispersion for an antiferromagnetic system, and the  minimum excitation energy induced by finite size effects. The ordering temperatures of some clusters (of size $L$ listed at the top of the figure) are shown by the vertical lines. For example, at $T$= 20 K clusters of linear size 64 \AA$ $ and smaller would order since not enough thermal energy would be available to start a disordering fluctuation with the maximum wavelength (minimum energy) $\lambda= 2L$.\label{finite}}
\end{center}
\end{figure}

When we cool down a chemically substituted system, some moments will be shielded while others survive. When we cool down further, we will find that groups of surviving moments will become isolated from the remainder of the moments because they find themselves surrounded by shielded moments. When that happens, the moments within such an isolated cluster will align with their neighbors because of finite size effects. These effects are illustrated in Fig. \ref{finite}. Essentially, it takes a certain amount of energy to keep moments from being ordered. Normally, above the ordering transition temperature, the thermal energy is sufficient to create so many excitations (spin-waves) that the moments will point in random directions. For small clusters, however, a finite amount of energy is required because (as quantum mechanics tells us) the disordering fluctuation cannot have a wavelength that is longer than twice the size of the cluster. Shorter wavelength excitations require more energy, and the energy requirements will be so large (see Fig. \ref{finite}) that they cannot be met by the available thermal energy close to 0 K. As a result, the moments on  small clusters are forced to order.\\

When the moments are aligned in a Kondo system, then the Kondo shielding mechanism will be rendered ineffective since this mechanism involves the spin flip of a conduction electron \cite{kondo}. This latter process is severely impeded in magnetically ordered surroundings. Therefore, once clusters split off from the infinite cluster, basic quantum mechanics dictates that the moments must order, thereby impeding the Kondo shielding mechanism. As a result, the moments of isolated clusters become impervious to being removed, and we end up with a restricted percolation model where moments can only be shielded (removed) on the magnetically disordered infinite cluster.\\

\begin{figure} [t]
\begin{center}
\includegraphics[width=380pt]{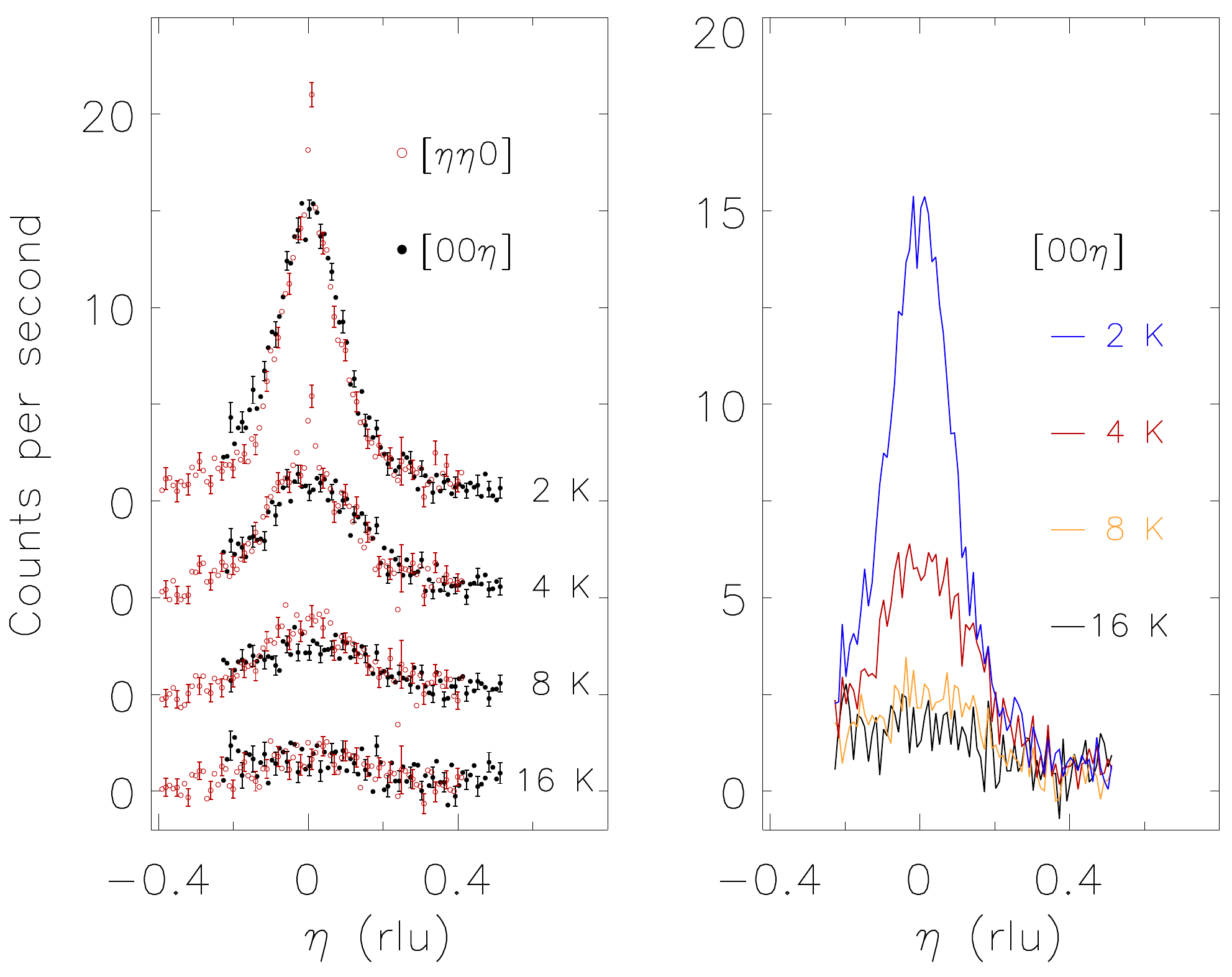}
\caption{Panel (a) displays the temperature dependence of the magnetic scattering associated with the formation of ordered clusters in Ce(Ru$_{0.246}$Fe$_{0.754}$)$_2$Ge$_2$, measured by means of neutron scattering \cite{montfrooij1}. One observes that the magnetic scattering starts to emerge at 16 K, and increases in intensity while narrowing in width in q-space when we cool down more. Since the inverse of this width is directly proportional to the distance over which moments are correlated, we see that (for all temperatures) there are just as many moments lined up along the c-direction (red symbols, momentum transfer along the (00$\eta$)-direction) as there are along another high symmetry direction (black symbols, momentum transfer along the ($\eta \eta$0)-direction). In this system, where the lengths of the a and c axes differ by a factor of 2.5, this must be indicative of clusters that are formed by random removal of magnetic moments. The curves have been offset along the vertical axis. Panel (b) displays the same data along the (00$\eta$)-direction, but now plotted on top of each other. Note the change in scale on the vertical axis. When the data are plotted this way one can observe that once clusters form and order, that this ordering remains intact upon cooling. When the temperature is lowered further, one observes that new (and larger, as implied by the narrower width) clusters form, and that the scattering by these clusters augments the scattering of the clusters already present. Panel (a) has been reproduced from W. Montfrooij {\it et al.}, Phys. Rev. B76, 052404 (2007).\label{qcp2}} 
\end{center}
\end{figure}

The quantum critical system Ce(Ru$_{0.246}$Fe$_{0.754}$)$_2$Ge$_2$-- that we alluded to before-- provides a good example of the shielding and cluster formation processes described above. When this system is cooled down, neutron scattering experiments \cite{montfrooij1,montfrooij2} reveal the formation of clusters and the ordering of the moments of the cluster members, as shown in Fig. \ref{qcp2}. Upon further cooling, additional clusters are formed, which also order the instant they separate from the infinite cluster. The experimental evidence for this is outlined in Fig. \ref{qcp2}. Following this direct observation of clusters in this antiferromagnetic system, the presence of magnetic clusters has since been inferred from uniform susceptibility measurements on various ferromagnetic quantum critical systems \cite{westerkamp, garciasoldevilla} and has been suggested for almost stoichiometric compounds\cite{lausberg}. Thus, we can take the presence of clusters in quantum critical compounds as an established fact.\\

In short, we have reasoned that restricted percolation should be relevant to chemically doped materials that undergo moment shielding upon cooling, while experiments have shown both the existence of clusters in such materials, as well as the moment alignment within isolated clusters. As such, it would seem highly likely that the response of such systems will reflect the presence of such clusters and the underlying percolative nature of the problem. For instance, the evolution of the specific heat with changing temperature should reflect the loss of entropy when the moments on a cluster are forced to order. The susceptibility should reflect whether isolated clusters have a net moment, and how easy it is for this net moment to be flipped. Close to the order-disorder transition, the temperature dependence of such quantities should be determined by the critical behavior inherent in the restricted percolative model. As such, studying this new set of models should help in understanding the unusual response of certain quantum critical systems, and perhaps even bring us closer to explaining why thus far these systems have eluded a full theoretical description.\\

We end this section with two remarks. First, the above strictly applies only to quantum critical systems that have been driven to criticality through chemical doping. We have argued elsewhere \cite{heitconf} that the restricted percolative model might also be relevant to stoichiometric systems, but that relevance requires a few more assumptions than the basic reasoning employed in this paper.\\

Second, it is not possible to directly relate the observed powerlaws in quantum critical systems to those of the restricted percolation model. The powerlaws in percolation are given as a function of site occupancy, whereas the experimental powerlaws are measured as a function of temperature. The missing link is the distribution function of exactly at what temperature a particular moment will be shielded. We will discuss such a distribution function \cite{bernal} in a forthcoming paper where we describe the observed response of Ce(Ru$_{0.246}$Fe$_{0.754}$)$_2$Ge$_2$ in terms of an underlying distribution of shielding temperatures in conjunction with the restricted percolation model.\\

In summary, we have discussed a restricted version of the standard percolation model. We have argued that this model bears direct relevance to the behavior of quantum critical system. We have derived relationships between the critical exponents in this restricted model and the exponents in the standard percolation model. We have verified by means of computer simulations that these derived relationships are correct, implying that our restricted model represents a new universality class.
\ack
This research is supported by the U.S. Department of Energy, Basic Energy Sciences, and the Division of Materials Sciences and Engineering under Grant No. DE-FG02-07ER46381. We are most thankful to Thomas Vojta for the many discussions regarding the validity of our restricted percolation model.
\section*{Appendix A}
\setcounter{section}{1}
In this appendix we derive eqn \ref{moments} under the assumption that whenever sites peel off from the infinite cluster that they all end up being part of a new single isolated cluster, rather than allowing for the possibility that multiple clusters peel off.\\

When a moment is removed from the infinite cluster, we can have two possible outcomes: either the infinite cluster loses one member, or a finite-sized cluster peels off from the infinite cluster. In the latter case the infinite cluster loses $n=-N\Delta P'$ members, and a new isolated cluster is created with $s=n-1$ members. Thus, at every percolation step $\Delta p'=-1/N$ that resulted in a change $\Delta P'(p')< 0$ in the infinite cluster membership probability, we find for the change in cluster moment distribution upon removing sites:
\begin{equation}
\Delta \sum s^{k} n_s(p')= (-N\Delta P'-1)^{k}/N,
\end{equation}
where the factor $1/N$ arises because $n_s(p')$ is the number of clusters with $s$ members {\it per lattice site}. Given that $-N\Delta P' -1$ equals $\Delta P'(p')/\Delta p'-1$ since $\Delta p' =-1/N$, this leads to
\begin{equation}
\frac{\Delta \sum s^{k} n_s(p')}{-\Delta p'}=\frac{-d\sum s^{k} n_s(p')}{dp'}=\left[ \frac{d P'(p')}{d p'}-1\right ]^{k}.
\end{equation}
Using $P'(p')= P_0'(p'-p_c')^{\beta'}+(p'-p'_c)$ and integrating the above equation we obtain
\begin{equation}
\sum s^{k} n_s(p') \sim (p'-p'_c)^{1-k+k\beta'}.
\end{equation}
Note that this equation only includes the singular part; the analytic part \cite{orbach}, if existent, has been omitted.

\end{document}